# Comparative Analysis of Cryptography Library in IoT

Uday Kumar
Tech Mahindra Limited
Chennai
India
udaykumar@techmahindra.com

*Tuhin Borgohain
Department of Instrumentation Engineering, Assam Engineering College
Guwahati
India
borgohain.tuhin@gmail.com

Sugata Sanyal
Corporate Technology Office,
Tata Consultancy Services
Mumbai
India
sugata.sanyal@tcs.com

*Corresponding Author

## ABSTRACT
The paper aims to do a survey along with a comparative analysis of the various cryptography libraries that are applicable in the field of Internet of Things (IoT). The first half of the paper briefly introduces the various cryptography libraries available in the field of cryptography along with a list of all the algorithms contained within the libraries. The second half of the paper deals with cryptography libraries specifically aimed for application in the field of Internet of Things. The various libraries and their performance analysis listed down in this paper are consolidated from various sources with the aim of providing a single comprehensive repository for reference to the various cryptography libraries and the comparative analysis of their features in IoT.

## 1. INTRODUCTION
The implementation of encryption and decryption in the field of cryptography provides a solid means of relaying messages to and fro between users without the added risk of the message being compromised to unwanted personnel. Such encryption-decryption operations are performed by various ways ([3], [7], [15]) through the use of specific set of algorithms. A cryptography library is a sort of repository of the various algorithms available for cryptographic purposes, which provides the added function of categorising the multitudes of algorithms into specific collections based on their performance capacities and functions.

In the field of IoT, microprocessors and embedded devices with low computational power plays the vital role of exchange of information using the internet infrastructure. Such constraints to computational capabilities and the necessity of secure exchange of information calls upon the need to implement algorithms specifically optimized to run in resource constrained environments. As such cryptography libraries aimed for use in microprocessors and embedded devices plays a very important role for providing the necessary security layers to IoT devices and securing up the overall IoT infrastructure.

## 2. OVERVIEW:
In this paper Section 3 will briefly introduce the various cryptography libraries available for encryption in general. It will also list all the encryption algorithms available in the various cryptography libraries. In section 4, we will discuss in details the various cryptography libraries in IoT. In section 5, we will do a comparative analysis amongst the various cryptography libraries discussed in section 4 based on their unique features. We conclude the paper in section 6.

## 3. CRYPTOGRAPHY LIBRARY
There exist numerous cryptography library encompassing multitudes of encryption algorithms which can be implemented for encryption of different messages in various fields. These cryptography libraries enable the implementation of various security measures ([11]) through the use of the containing algorithms. Some of the most prominent cryptography library ([5]) along with their encryption algorithms is listed below:

i. Borzoi: The "borZoi" cryptography library implements an algorithm based on elliptic curves (as such known as Elliptic Curve Cryptography Library) ([4], [9], [10], [14], [36]). It implements the following algorithms which ranges over a finite field bearing a characteristic 2 (GF2m) ([1]):
    a.    ECDSA (Elliptic Curve Digital Signature Algorithm)
    b.    Elliptic Curve Diffie-Hellman Key Agreement Scheme
    c.    ECIES (Elliptic Curve Integrated Encryption Scheme)

borZoi is also implemented with AES Symmetric encryption scheme and one other algorithm to produce SHA-1, its digital signature which are as follows ([1]) :
    a.    AES (Rijndael) Symmetric Encryption Scheme
    b.    SHA-1 hash algorithm

ii. Crypto++ : Written in C++, this cryptography library implements various algorithms ranging from authenticated encryption schemes (like GCM, CCM etc.) to algorithms based on elliptic curves (like ECDSA, ECNR etc) ([13]). The various algorithms implemented by Crypto++ are as follows ([2]):
    a.    GCM, CCM, EAX

  b. AES (Rijndael), RC6, MARS, CAST-256, Twofish, Serpent
  c. Panama, Sosemanuk, Salsa20, XSalsa20
  d. IDEA, Triple-DES, Camellia, SEED, XTEA, Skipjack, SHACAL-2, RC5, Blowfish
  e. ECB, CBC, CTS, CFB, OFB, CTR
  f. VMAC, HMAC, CBC-MAC, GMAC, Two-Track-MAC
  g. SHA-1, SHA-2, SHA-3, WHIRLPOOL, Tiger, RIPEMD-128, RIPEMD-256, RIPEMD-160, RIPEMD-320
  h. ECDSA, ECNR, ECIES, ECMQV, ECDH
  i. MD2, MD4, MD5, Panama Hash, Square, GOST, SAFER, SEAL 3.0, DES, ARC4, DESX, RC2, 3-WAY, WAKE-OFB, CAST-128, SHARK
  j. Diffie-Hellman, XTR-DH, DH2, MQV, LUCDIF
  k. PKCS#1 v2.0, OAEP, PSS, IEEE P1363 EMSA2-EMSA5, PSSR
  l. ESIGN, LUC, RSA, DSA, ElGamal, RW, NR, DLIES

 iii. Libmcrypt: The "libmcrypt" cryptography library provides encryption of data and is thread safe. This specific library contains a set of encryption algorithms and modes which are modular in nature. This nature allows algorithms and the encryption modes to operate in a much efficient manner. The various algorithms contained within the framework of this library are tabulated in Table 1:

| xTEA | CAST-128 | CAST-256 | DES | 3DES | GOST | SKIPJACK |
|---|---|---|---|---|---|---|
| 3-WAY | BLOWFISH | TWOFISH | WAKE | PANAMA | MARS | LOKI97 |
| RC2 | RC6 | ARCFOUR | RIJNDAEL | CBC | ECB | SAFER |
| SAFER+ | SAFER K-64 | SAFER K-128 | SAFER SK-64 | SAFER SK-128 | ENIGMA | IDEA |
| SERPENT | STREAM | CFB | OFB | nOFB | nCFB | CTR |

Table 1: Algorithms in Libmcrpyt library

 iv. Botan (formerly known as OpenCL): This cryptography library is written in C++ and licensed under BSD-2 ([23], [28]). It was later implemented with a "Card Verifiable Certificate" for ePassports and this modified version of Botan was named "InSiTO". This library contains a number of encryption formats, algorithms and protocols which are tabulated in Table 2:

| TLS | SSL | PKCS | PKCS #3 | PKCS #5 (v1.5/v2.0) |
|---|---|---|---|---|
| RSA | DSA | X.509 CRLs | Parts of 1363 | Diffie-Hellman |

Table 2: Algorithms in Botan library

 v. Libgcrypt: Written in C language, the "libgcrypt" is a multi-platform cryptography library licensed under GNU Lesser General Public License GNU General Public License ([32]). It features a multiple precision arithmetic implementation and entropy gathering utility ([37]). The cryptography algorithms in this library are tabulated in Table 3:

| IDEA | 3DES | SERPENT (128 bits) | SERPENT (192 bits) | SERPENT (256 bits) | CAST5 | BLOWFISH | AES 128 |
|---|---|---|---|---|---|---|---|
| AES 192 | AES 256 | TWOFISH (128 bits) | TWOFISH (256 bits) | ARCFOUR | DES | Ron's Cipher 2 (40 bits) | Ron's Cipher 2 (128 bits) |
| SEED | Camellia (128 bits) | Camellia (192 bits) | Camellia (256 bits) | Salsa20 | Salsa20/12 | GOST 28147-89 | STREAM |
| GCM | CCM | RFC 3394 | CFB | CBC | ECB | OFB | CTR |
| RSA | DSA | ElGamal | ECDSA | EdDSA | CMAC | GMAC | HMAC |
| SHA-1 | TIGER | RIPEMD-160 | MD4 | MD5 | TIGER/192 | TIGER2 | SHA-224 |
| Whirlpool | GOST R 34.11-2012 (256 bits) | GOST R 34.11-2012 (512 bits) | SHA-256 | SHA-384 | SHA-512 | ISO 3309 | RFC 1510 |
| RFC 2440 | GOST R 34.11- | RFC 4880 | PBKDF2 | SCRYPT | | | |

Table 3: Algorithms in Libgcrypt library

 vi. Bouncy Castle: This particular cryptography library is written in Java and C# ([41]). Designed mainly for use in devices with low computational memory, this library contains the algorithms listed in Table 4:

| PKCS#10 | DANE | DVCS | OCSP | DTLS | OpenPGP | CRMF |
|---|---|---|---|---|---|---|
| CMP | TSP | TLS | PKCS#12 | CMS | S/MIME | DTLS |

Table 4: Algorithms in Bouncy Castle library

 vii. Cryptlib: The "cryptlib" cryptography library is a library of algorithms which provides security to communication and information exchange. Its simple interface makes it very user-friendly and its layered structure (the lower layers each providing a layer of abstraction, the higher layers covering up the details of implementation of the algorithms) makes up the whole library very secure and impermeable to intrusion to a very high degree. The various algorithms within this library are tabulated in Table 5:

| SSL | TLS | SSH | S/MIME | OpenPGP | CMP | SCEP | RTCS |
|---|---|---|---|---|---|---|---|
| OCSP | X.509v1 | SET | Microsoft AuthentiCode | RPKI | SigG | Identrus | PKCS #7 |
| RTCS | OCSP | CA | X.509v3 | | | | |

Table 5: Algorithms in Cryptlib library

 viii. Catacomb: Written using gcc, this cryptography library contains a set of cryptographic primitives and used in Linux operating systems ([9]). Some of the most prominent categories of algorithms within this library out of its many other are as shown in Table 6:

| BLOCK Cipher | HASH functions | Multi-precision Maths Library | Public Key Algorithms |
|---|---|---|---|

Table 6: Categories of algorithms in the Catacomb library

 ix. Cryptix: The "Cryptix" (*say* Cx) cryptography library was made to provide a library of cryptographic algorithms to the Java platform as there were a number of issues regarding adoption of cryptography in Java ([22]). With the removal of export controls on cryptography, the use of "Cryptix" (last active development was in 2005) declined with the increasing availability of other more secure cryptography libraries. The list of algorithms under this library are shown in Tale 7:

| Cx OpenPGP | Cx Perl | Cx Perl PGP | Cx JCE | Cx SASL | Cx ASN.1 |
|---|---|---|---|---|---|
| Cx v3.1.3 | Cx v3.1.3 PGP | Cx v3.2.0 | Cx v3.2.0 PGP | Cx AES Kit | Cx Elliptix |

Table 7: Algorithms in Cryptix library

 x. Flexiprovider: This cryptography library is built for use in encryption of any application built upon the JCA (Java

Cryptography Architecture) ([39]). This encryption toolkit is supported by CoreProvider (containing algorithms like PKCS #1, 3DES etc.), ECProvider (which contains algorithms based on elliptic curve such ECDH key agreement scheme, ECDSA etc.), PQCProvider (Contains the McEliece cryptosystem in four variants (CFS signature scheme etc.) and NFProvider (contains IQRDSA, IQDSA, IQGQ etc.).

xi. LibTomCrypt: The "LibTomCrypt" cryptography library is an open source library of cryptographic primitives ([20]).

xii. MatrixSSL: The "MatrixSSL" cryptography library is designed for devices and application with smaller footprint. An implementation of embedded SSL and TLS, it contains various symmetric key and public key algorithms. Some popular algorithms included in this library are given in Table 8:

| RSA | Diffie-Hellman | Elliptic Curve Cryptography | AES |
|---|---|---|---|
| AES-GCM | SEED | ARC4 | 3DES |

Table 8: Algorithms in MatrixSSL

xiii. MIRACL: Multiprecision Integer and Rational Arithmetic C Library (MIRACL) is a cryptography library designed for use in constrained environment in terms of size and computational power ([38]).

xiv. Mozilla's NSS: NSS (Network Security Services) cryptography library facilitates the encryption in server-based applications. It mainly supports the following security algorithms listed in Table 9 for use in server applications:

| SSL | S/MIME | TLS | PKCS #11 |
|---|---|---|---|

Table 9: Security algorithms in NSS

xv. OpenPGP: This cryptography library is an open source variant of PGP (Pretty Good Privacy) which is used for securing the privacy of end-users and levelling up the security of communication systems by implementation of authentication methods through the use of PGP ([16], [18]).

xvi. OpenSSL: Written in C language, the "OpenSSL" is a multi-platform library of cryptographic algorithms and functions ([40]). It is an open source library licensed under Apache License 1.0 and 4-clause BSD License. It implements the various SSL protocols and TLS protocols. The various algorithms implemented by "OpenSSL" are tabulated in Table 10 as follows:

| 3DES | RC2 | RC4 | RC5 | BLOWFISH |
|---|---|---|---|---|
| Camellia | AES | SEED | CAST-128 | IDEA |
| DES | MD2 | MD4 | MD5 | SHA-1 |
| SHA-2 | RIPEMD-160 | MDC-2 | DSA | RSA |
| Diffie-Hellman | Elliptic Curve | GOST R 34.11-94 | GOST R 34.10-2001 | GOST 28147-89 |

Table 10: Algorithms implemented by OpenSSL

xvii. Nettle: This is a low-level, multi-platform cryptography library licensed under GNU Lesser General Public License ([17]). The various algorithms within this cryptography library are shown in Table 11:

| AES BLOCK Cipher | RC4 | RC2 | BLOWFISH | Camellia | CAST-128 | DES | 3DES |
|---|---|---|---|---|---|---|---|
| ChaCha STREAM Cipher | GOSTHASH94 | RSA | DSA | ECDSA | TWOFISH | SHA-3 | SHA224 |
| SHA256 | SHA384 | SHA512 | SHA-1 | SERPENT | Salsa20 | RIPEMD160 | UMAC |
| POLY1305 | PBKDF2 | MD2 | MD4 | MD5 | Yarrow pRNG | | |

Table 11: Algorithms in Nettle

## 4. CRYPTOGRAPHY LIBRARIES IN IoT

### 4.1 WolfSSL (formerly known as CyaSSL):
Written in ANSI C, the "wolfSSL" cryptography library, due to its small footprint size and low runtime memory, is aimed to be used in embedded devices, RTOS and environments facing constraints in computational resources ([30], [33]). This library supports the development of cross-platform algorithms and houses a large number of algorithms. Moreover it features the generation of Key and Certificates. "wolfSSL" is licensed under GNU General Public License GPLv2.

"wolfSSL" contains the following categories of algorithms to be used for cryptographic purposes which are shown in Table 12:

| CATEGORY | ALGORITHMS | | | | | | |
|---|---|---|---|---|---|---|---|
| wolfCrypt | RSA | DSS | SHA-1 | SHA-2 | ECC | BLAKE2 | Poly1305 |
| | Diffie-Hellman | EDH | DES | 3DES | GCM | CCM | CTR |
| | CBC | Camellia | ARC4 | HC-128 | ChaCha20 | Random Number Generation | Rabbit |
| | MD2 | MD4 | MD5 | | | | |
| NTRU | AES-256 | RC4 | HC-128 | | | | |

Table 12: Algorithm library of wolfSSL

### 4.2 AvrCryptoLib:
Licensed under GPLv3, the "Avr-Crypto-Lib" cryptography library has the implementation of its encryption algorithms in the AVR 8-bit microcontrollers ([34]). As with all the rest of the cryptography library aimed to be used in the field of IoT, the "Avr-Crypto-Lib" is optimized for resource-constrained environments in regards to available computational memory and size.

"Avr-Crypto-Lib" contains vast number algorithms which are categorised in Table 13:

| CATEGORY | ALGORITHMS | | | | | | |
|---|---|---|---|---|---|---|---|
| STREAM Cipher | ARC4 | Trivium | Mugi | Grain | Mickey | | |
| BLOCK Cipher | AES | XTEA | CAST5 | Camellia | Threefish-256 | Threefish-512 | Threefish-1024 | SEED |
| | SERPENT | SHABEA | Present | SKIPJACK | Noekeon | RC5 | RC6 | DES |
| | EDE-DES | 3DES | | | | | |
| HASH Functions | BLAKE | Twister | Shabal | Skein | SHA-1 | SHA-256 | MD5 | Grost1 |
| | BlueMidnightWish | | | | | | |

Table 13: Algorithm library of Avr-Crypto-Lib

Besides the above mentioned algorithms, "Avr-Crypto-Lib" also provides MAC functions and Pseudo Random Number Generators (PRNGs).

### 4.3 WiseLib:
Written in C++, the "Wiselib" cryptography library is targeted to be used in networked embedded devices ([12], [35]). Using "Wiselib", an individual can compile algorithms for various platforms like Contiki, iSense, Shawn (a simulator of sensor network) etc. using its various in-house algorithms like routing algorithms, localization algorithms etc. The use of template similar to Boost and CGAL facilitates highly efficient compilations of the various generic and platform independent codes written for various platforms.

**4.4 TinyECC:** TinyECC is an Elliptic Curve Cryptography based library which can perform ECC-based PKC operations ([25], [29]). Some of the most prominent features of TinyECC are:

a. Provision of ECDSA, a digital signature
b. ECIES, a scheme for encryption of public key
c. ECDH, a protocol for key exchange

**4.5 RelicToolKit:** Licensed under LGPL v2.1 (and above), the "RELIC toolkit" cryptography library is an efficient and flexible meta-toolkit ([21]). The main use of the "RELIC toolkit" is in its ability to be used for construction of custom cryptographic toolkits.

The various algorithms implemented by the "RELIC toolkit" are as follows:
- Multiple-precision integer arithmetic
- Bilinear maps and extensions fields relate to bilinear maps
- Elliptic curves:
  - Over prime fields
  - Over binary fields
- Prime and Binary field arithmetic
- Cryptographic protocols

The various cryptographic protocols implemented by RELIC are tabulated in Table 14:

| ECDSA | RSA | ECIES | ECSS | ECMQV |
|---|---|---|---|---|
| Rabin | Sakai-Ohgishi-Kasahara ID-based authenticated key agreement | Boneh-Lynn-Schacham short signature | Boneh-Boyen short signature | Paillier and Benaloh homomorphic encryption systems |

Table 14: Cryptographic protocols of RELIC toolkit

# 5. COMPARATIVE ANALYSIS OF THE CRYPTOGRAPHY LIBRARIES IN IoT

**5.1 WolfSSL:** WolfSSL includes OpenSSL compatibility layer along with support for OCSP and CRL which are used for validating certificates. Its runtime memory usage is between 1-36 kB. Sporting a very simple API, this library supports zlib compression, IPv4 and IPv6 along with integration of MySQL ([30]).

**5.2 AvrCryptoLib:** This cryptography library performs modular exponentiation using C-interfaces in AVR 8-bit assembly language. This leads to reduction in execution time of this cryptography library. Moreover this library allows direct access to keys through storage of these keys in the flash memory which results in efficient consumption of SRAM.

**5.3 WiseLib:** Implementing elliptic curve over prime fields only, this library shuns away from incorporation of optimizations of assembly level for making its codes platform independent.

**5.4 TinyECC:** Though mainly made for running in devices operating on TinyOS, TinyECC can be implemented in devices other than TinyOS-dependent devices as the library can be ported to C99 through manual alteration of code parts or through the usage of tool-chains. This library also implements curves over prime fields only and includes sliding windows ([43]) and Barrett reduction ([44]) for the purpose of verification.

**5.5 RelicToolKit:** The inclusion of multiple integer arithmetic makes its compilation of this library easy for a wide variety of platforms. RelicToolKit provides high level of customization in terms of:

*5.5.1* Building and inclusion of desired components only for usage in desired platforms.

*5.5.2* Desired selection of various mathematical optimizations for optimum performance of the toolkit in a specific platform.

# 6. CONCLUSION

From the above comparative analysis of the various features available in the different cryptography library the foremost conclusion is that not one of the cryptography libraries in the IoT environment can be considered as a universal library due to their varying features and optimizations made for different specific platforms. This results in the non-existence of a single universal standard library that can be applicable to all IoT devices around us. Moreover each library contains a specific set of features unique to them and optimized for the platform where these are applicable. And the use of the above cryptography libraries along with the adoption of various security measures that can be adopted in various communication modes ([24], [42]) and implementation of intrusion detection systems and schemes ([19], [27], [31]) will lead to a more secure and reliable IoT infrastructure for wide adoption of its devices by the masses.